\documentclass[aps, prl,twocolumn]{revtex4}
\usepackage{graphicx}
\usepackage{dcolumn}
\usepackage{bm}
\usepackage{epstopdf}
\usepackage{color}
\begin{document}

\title{Wavelength tunable soliton rains in a nanotube-mode locked Tm-doped fiber laser}
\author{B. Fu$^1$$^,$$^2$$^,$$^3$, D. Popa$^1$, Z. Zhao$^1$, S. A. Hussain$^1$, E. Flahaut$^4$, T. Hasan$^1$, A. C. Ferrari$^1$}
\affiliation{$^1$Cambridge Graphene Centre, University of Cambridge, Cambridge CB3 0FA, UK}
\affiliation{$^2$Beijing Advanced Innovation Center for Big Data-Based Precision Medicine, Beihang University, Beijing 100191, China}
\affiliation{$^3$School of Instrument Science and Opto-Electronic Engineering, Beihang University, Beijing, 100191, China}
\affiliation{$^4$University Paul Sabatier, CIRIMAT/LCMIE, CNRS UMR 5085, Toulouse, France}

\begin{abstract}
We report soliton rains in a tunable Tm-doped fiber laser mode locked by carbon nanotubes. We also detect their second- and third-harmonic. We achieve a tunability of over 56nm, from 1877 to 1933nm, by introducing a polarization-maintaining isolator and two in-line polarization controllers. This makes our system promising as a tunable filter for ultrafast spectroscopy.
\end{abstract}
\maketitle
Soliton rains (SR) in mode-locked fiber lasers are drifting pulses generated from the continuous wave (CW) background noise\cite{Chouli09oe,Chouli10pra} or the condensed soliton phase (CP) (i.e., a static pulse instead of drifting ones)\cite{Bao13ol,Huang14lpl,Singh16ptl,Meng12oe}. These drift towards the next adjacent CP spontaneously\cite{Chouli09oe,Chouli10pra,Bao13ol,Huang14lpl,Singh16ptl,Meng12oe,Niang14apb,Xu15lpl}. They were first reported in Refs.\cite{Chouli09oe,Chouli10pra}, and since then in many other mode-locked fiber lasers\cite{Bao13ol,Huang14lpl,Singh16ptl,Meng12oe,Niang14apb,Xu15lpl}. SR are characterized by 3 properties: drift, CP and CW background noise\cite{Chouli09oe,Chouli10pra}. SR were first reported in Er-doped fiber lasers\cite{Chouli09oe,Chouli10pra}, generated from the CW background noise and drifting towards the CP, triggered by the injection of an external CW laser\cite{Chouli09oe,Chouli10pra}. SR were also generated from the CP in a normal dispersion Yb-doped fiber laser with a dual-filter effect\cite{Bao13ol}, providing insights on the dissipative soliton SR dynamics\cite{Bao13ol}. SR have been studied at 1$\mu$m\cite{Bao13ol,Huang14lpl,Singh16ptl} and 1.5$\mu$m\cite{Chouli09oe,Chouli10pra,Meng12oe,Niang14apb}, with one report at 2$\mu$m based on a nonlinear amplified loop mirror cavity\cite{Xu15lpl}. However, Ref.\cite{Xu15lpl} exploited a two-ring cavity design, with a rather long cavity (116m) compared to standard ($\sim$10m) soliton fiber lasers\cite{Chouli09oe,Chouli10pra,Bao13ol,Huang14lpl,Singh16ptl,Meng12oe,Niang14apb}, and no wavelength tunability\cite{Xu15lpl}. Therefore, to better understand the soliton dynamics in different wavelength regions (e.g., the 2$\mu$m standard soliton fiber lasers\cite{Zhang12oe}), it is necessary to find simple, wavelength tunable methods able to support SR. Tm-doped fiber lasers operating in the 2$\mu$m region an eye-safe band because liquid water (main constituent of human tissues) absorbs strongly\cite{Nelson95apl}, have potential applications in molecular spectroscopy\cite{Walsh09lp}, optical communications\cite{Li13oe}, medicine\cite{Fried05je} and light detection and ranging (LIDAR)\cite{Henderson93rs}.

Here, we report soliton rains in a broadband wavelength tunable Tm-doped fiber laser mode-locked by a carbon nanotube (CNTs) saturable absorber (SA). The second- and third-harmonic SRs are also observed. The output wavelength can be tuned from 1877 to 1933nm by tuning two in-line polarization controllers, making this promising as tunable filter for ultrafast spectroscopy.

CNTs can be used as SAs for mode-locking\cite{Wang08nn,Going12pe,Kieu10ptl,Hasan14an,Meng17sr,Zhang15apl,Popa12apl,Hasan09am,Martinez13np}, as they provide high modulation depths ($\sim$17\%\cite{Popa12apl}), preferred for mode-locked fiber lasers\cite{Wang08nn,Going12pe,Kieu10ptl,Hasan14an,Meng17sr,Zhang15apl,Popa12apl,Hasan09am,Martinez13np} typically operating with higher gain and cavity losses than their solid-state counterparts\cite{Okhotnikov_fl}, which is much larger than what is usually reported for single layer graphene devices\cite{Martinez13np,Zaugg13oe,Cho11ol} (e.g., $\sim$0.5\% in Ref.\cite{Cho11ol}). Here we use solution processed CNTs to fabricate the SAs\cite{Hasan09am}. 0.03wt\% purified laser ablation single-wall CNTs\cite{Lebedkin02c,Hennrich03pccp} and 0.02wt\% of purified catalytic CVD grown double-wall CNTs (DWNTs)\cite{Osswald06cm} are ultrasonicated separately for an hour with 0.7wt\% sodium-carboxymethylcellulose (Na-CMC) polymer using a tip sonicator at 100W output power and 10$^{\circ}$C for 4 hours. The dispersions are then ultra-centrifuged. The topmost 60\% is decanted, then mixed and ultrasonicated again for 30mins, then drop-cast and the solvent evaporated in a desiccator, resulting in a$\sim$30$\mu$m thick CNT-CMC composite. The mean inner and outer diameters of the DWNTs are$\sim$1.1 and 1.8nm\cite{Hasan14an}, resulting in absorption peaks at$\sim$1.1 and 1.9$\mu$m\cite{Hasan14an,Liu12nn}.
\begin{figure}
\centerline{\includegraphics[width=80mm]{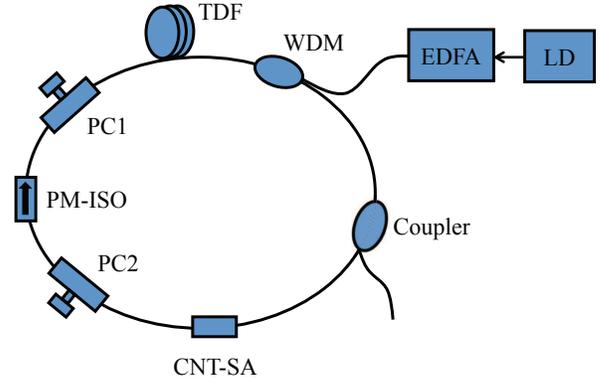}}
\caption{Fiber laser set up. LD: laser diode, EDFA: Erbium-doped fiber amplifier, WDM: wavelength-division multiplexer, TDF: Tm-doped fiber, PC: polarization controller, PM-ISO: polarization-maintaining isolator, CNT-SA: carbon nanotube saturable absorber.}
\label{fig1}
\end{figure}

A schematic of our Tm-doped fiber laser setup is shown in Fig.\ref{fig1}. It consists of a 2.9m Tm-doped fiber (TDF, SCF-TM-8/125, with group velocity dispersion (GVD)$\sim$69ps$^2$/km at 1900nm), a wavelength-division multiplexer (WDM) to couple the pump source (1560nm laser diode (LD) amplified by an Er-doped fiber amplifier, EDFA) into the TDF, a polarization-maintaining isolator (PM-ISO, 0.81m PM 1550 fiber pigtail) to ensure unidirectional operation, two in-line polarization controllers (PCs) to adjust cavity polarization, and an optical coupler with a 20\% output. The CNT-SA is sandwiched between the two fiber ferrules. The rest of the cavity is a standard single-mode fiber (SMF-28, with GVD$\sim$67ps$^2$/km at 1900nm), with length$\sim$11.7m. The total dispersion is$\sim$-0.79 ps$^2$. The output is monitored by an optical spectrum analyzer, an oscilloscope via an extended InGaAs photodetector, an autocorrelator, a spectrum analyzer, and a power meter.
\begin{figure}
\centerline{\includegraphics[width=90mm]{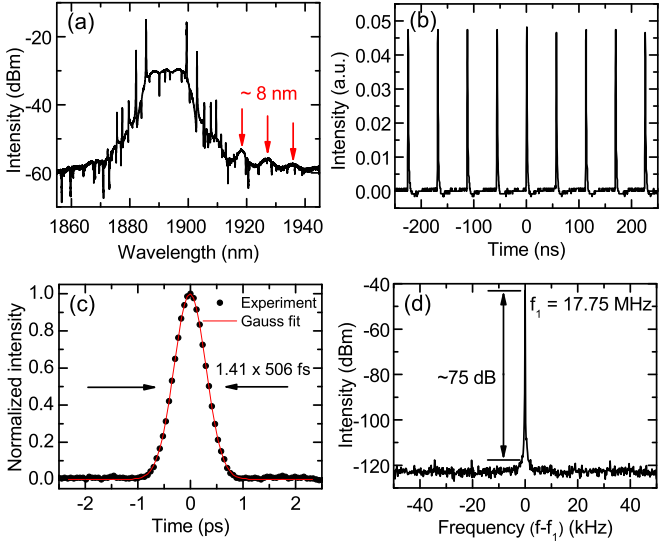}}
\caption{(a) Optical spectrum. (b) Temporal waveform. (c) Autocorrelation trace. (d) RF spectrum.}
\label{fig2}
\end{figure}

Mode-locking self-starts when the pump power reaches$\sim$500mW. The output power is 16.3mW and the pulse energy is 0.92nJ. As shown in Fig.\ref{fig2}(a), the central wavelength is 1893nm with full width at half maximum (FWHM)$\sim$14.5nm. The arrows in Fig.\ref{fig2}(a) indicate the 8nm wavelength spacing caused by the Lyot birefringence filter effect\cite{Ozgoren10ol}, attributed to the two PCs and the PM-ISO. This can be calculated as $\Delta\lambda=\lambda^2/(L\Delta n)$\cite{Ozgoren10ol}, where \textit{L} is the length of the PM fiber and $\Delta n$ is the modal birefringence. The sharp dips in the spectrum are attributed to the intrinsic absorption peaks and molecular resonances of the Tm-doped fiber\cite{Stark03oc,Sobon15oe}. Fig.\ref{fig2}(b) shows the temporal waveform at the fundamental repetition rate (FRR)$\sim$17.75MHz, corresponding to the cavity length$\sim$11.7m. Fig.\ref{fig2}(c) plots the autocorrelation trace of the pulses. A good fit is obtained by using a Gauss profile, with FWHM$\sim$506fs. The time-bandwidth product (TBP) is 0.614, larger than the theoretical value (0.441) for transform-limited of Gauss-shaped pulse\cite{Agrawal_anfo}, indicating minor chirping\cite{Agrawal_anfo}. The signal-to-noise ratio (SNR) of the radio-frequency (RF) spectrum with 10Hz resolution [Fig.\ref{fig2}(d)] is 75dB at the FRR, which indicates stable mode-locking\cite{Linde86apb}.
\begin{figure}
\centerline{\includegraphics[width=90mm]{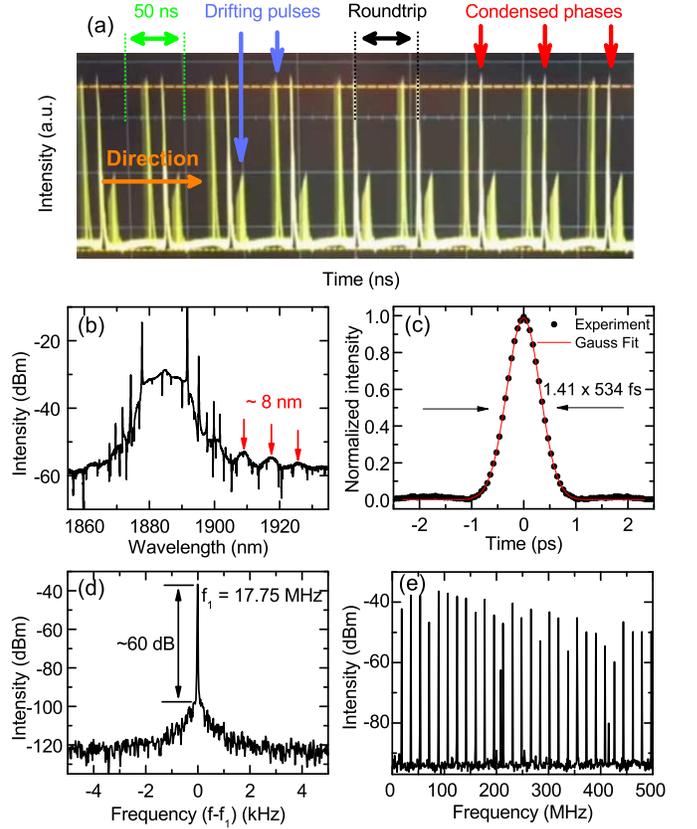}}
\caption{(a) Oscilloscope trace of SR. (b) Optical spectrum. (c) Autocorrelation trace. (d) RF spectrum. (e) RF spectrum with 500MHz span.}
\label{fig3}
\end{figure}

By increasing the pump power and adjusting the PCs, we achieve SR. Fig.\ref{fig3}(a) plots the SR oscilloscope trace when we increase the pump power to$\sim$650mW. The output power is$\sim$43.3mW. The red arrows indicate CPs that radiate the drifting pulses (blue arrows) from left to right. During the SR process, the drifting pulses spontaneously start from one CP and vanish close to the next CP\cite{Bao13ol,Huang14lpl,Singh16ptl,Meng12oe,Niang14apb,Xu15lpl}. The spacing of two adjacent CPs is the FRR\cite{Chouli09oe,Chouli10pra}. The central wavelength of the spectrum [Fig.\ref{fig3}(b)] is 1884nm with FWHM$\sim$10.4nm. Fig.\ref{fig3}(c) shows the autocorrelation trace of the soliton pulses. A good fit is obtained by using a Gauss profile, with FWHM$\sim$534fs. The TBP is 0.469, close to the theoretical value\cite{Agrawal_anfo}. The SNR of the RF spectrum with 10Hz resolution [Fig.\ref{fig3}(d)] is 60dB at the FRR, smaller than the normal mode-locked SNR ($\sim$75dB) caused by the soliton flow (i.e., SR), which decreases the SNR\cite{Linde86apb}. We also measure the RF spectrum with 500MHz span, Fig.\ref{fig3}(e), and attribute its fluctuation to the soliton flow\cite{Linde86apb}.
\begin{figure}
\centerline{\includegraphics[width=90mm]{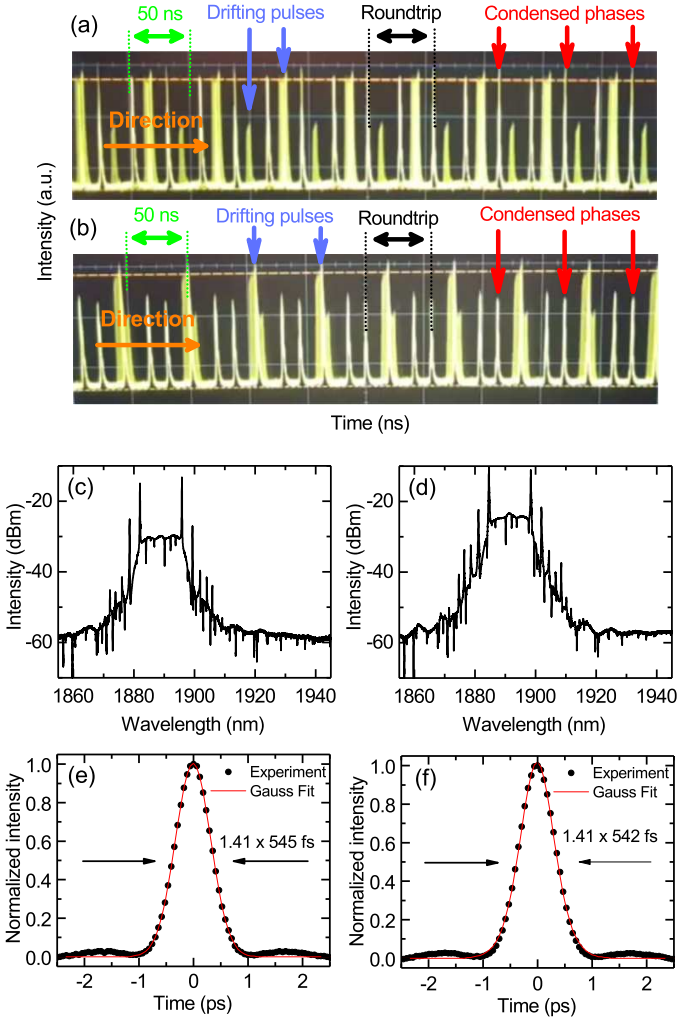}}
\caption{Oscilloscope traces of (a) second and (b) third-harmonic soliton rains. Spectra of (c) second- and (d) third-harmonic SR. Autocorrelation traces of (e) second- and (f) third-harmonic SR.}
\label{fig4}
\end{figure}

SR in our laser generate from CP and drift towards the next CP. During the process, the amplitudes of the drifting soliton pulses enlarge before they reach CP. This is different from the SR observed in Refs.\cite{Chouli09oe,Chouli10pra,Bao13ol,Huang14lpl,Singh16ptl,Meng12oe,Niang14apb,Xu15lpl}, where tens of drifting pulses within a roundtrip time equally distribute their energy and aggregate during the flow of the soliton pulses\cite{Chouli09oe,Chouli10pra,Bao13ol,Huang14lpl,Singh16ptl,Meng12oe,Niang14apb,Xu15lpl}. Our laser has only two drifting pulses in a roundtrip [Fig.\ref{fig3}(a)]. Thus, they can aggregate energy before reaching the next CP, and subsequently release energy, i.e., radiate SR, in dynamic equilibrium. SR lasts tens of minutes if one does not operate the PC.
\begin{figure}
\centerline{\includegraphics[width=90mm]{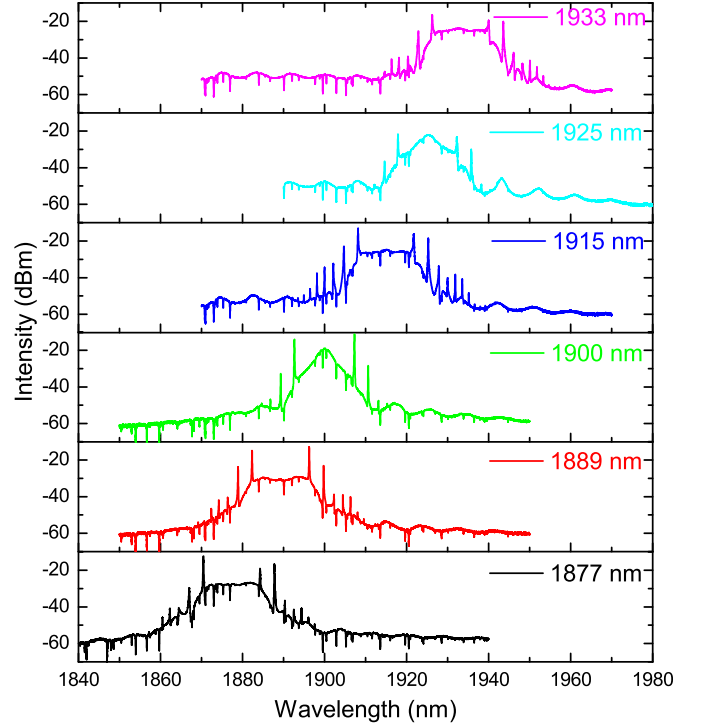}}
\caption{Wavelength tunable spectra.}
\label{fig5}
\end{figure}

By further increasing the pump power and adjusting the PCs, we observe the second- and third-harmonic SR, Figs.\ref{fig4}(a,b). Compared to the FRR SR, the CP and the drifting direction of the pulses are unchanged. The drifting pulses still arise from CP and drift from left to right for a roundtrip route, with the amplitude enhanced when they pass the second- and third-harmonic, due to the superposition of the harmonics and the drifting pulses. The drift speed can be changed by adjusting the PCs, since SR are sensitive to the cavity polarization\cite{Huang14lpl,Niang14apb,Xu15lpl}. Figs.\ref{fig4}(c,d) plot the output wavelengths corresponding to the second- and third-harmonic SR. The central wavelengths are 1889 and 1892nm, with FWHM$\sim$14.2 and 14.7nm, respectively. Figs.\ref{fig4}(e,f) show the autocorrelation traces of the second- and third-harmonic SR. The FWHM is 545 and 542fs, by using Gauss pulse profile, with TBP$\sim$0.651 and 0.668, respectively, larger than the theoretical value for the transform-limited of Gauss-shaped pulse\cite{Agrawal_anfo}, indicating chirping.

For SR generation, the drifting pulses move faster than the CP. This means that the SR have different group velocity than the CP, corresponding to different central wavelengths\cite{Chouli09oe,Chouli10pra,Bao13ol}. Thus, the SR spectra always show modulations (i.e., extra optical sidebands in addition to Kelly sidebands\cite{Kelly92el}) because of the interactions between different SRs and CP central wavelengths\cite{Chouli09oe,Chouli10pra,Bao13ol,Huang14lpl,Singh16ptl,Meng12oe,Niang14apb,Xu15lpl}. In our laser, the PM-ISO along with the two PCs act as a Lyot birefringence filter\cite{Ozgoren10ol}, enhancing spectral modulation\cite{Villanueva12ol}, which is beneficial for SR generation\cite{Bao13ol}.

By adjusting the PCs, we obtain SR tunability from 1877 to 1933nm as shown in Fig.\ref{fig5}. As indicated in Table \ref{table}, the central wavelengths are 1877, 1889, 1900, 1915, 1925 and 1933nm with FWHM$\sim$14.6, 15.2, 3.1, 13.9, 3.5 and 14.1nm, respectively. We assume Gaussian pulses for 1877, 1889, 1915 and 1933nm (as confirmed by spectral modulation\cite{Ozgoren10ol,Villanueva12ol}), and sech$^2$ for 1900 and 1925nm (as confirmed by the Kelly sidebands\cite{Kelly92el}). The pulse duration is 527, 523, 1284, 543, 1233 and 534fs, corresponding to TBP$\sim$0.655, 0.668, 0.331, 0.617, 0.349 and 0.605, respectively, indicating chirping for the Gaussian pulses and almost transform-limited behavior for the sech$^2$ ones (0.315 holding for a transform-limited pulse\cite{Agrawal_anfo}). The different wavelength and pulse profiles are attributed to the birefringence filter effect\cite{Ozgoren10ol,Villanueva12ol}.
\begin{table}
\caption{Output properties of wavelength tunable spectra}
\begin{tabular}{cccc}
\hline
Wavelength & FWHM & Pulse Duration & TBP \\
(nm) & (nm) & (fs) &  \\
\hline
1877 & 14.6 & 527 & 0.655 \\
1889 & 15.2 & 523 & 0.668 \\
1900 & 3.1 & 1284 & 0.331 \\
1915 & 13.9 & 543 & 0.617 \\
1925 & 3.5 & 1233 & 0.349 \\
1933 & 14.1 & 534 & 0.605 \\
\hline
\end{tabular}
  \label{table}
\end{table}

In summary, we reported the fundamental repetition rate, second- and third-harmonic soliton rains in a wavelength tunable (1877-1933nm) Tm-doped fiber laser mode-locked by carbon nanotubes. The pulse duration of the fundamental, second- and third-harmonic SR is 534, 545 and 542fs, respectively. The condensed soliton phase and the drifting direction of the pulses are unchanged during the process. This makes our system promising as a tunable filter for ultrafast spectroscopy.

We acknowledge funding from ERC Grant Hetero2D, EPSRC Grants EP/L016087/1, EP/K017144/1, EP/K01711X/1 and the China Scholarship Council.

\end{document}